\documentclass[11pt,onecolumn,floatfix,superscriptaddress]{revtex4}

\usepackage{graphicx}                   
\usepackage{epsfig}
\usepackage{amsmath,amssymb}            
              
\usepackage{epstopdf}
\usepackage{placeins}
\usepackage{upgreek}

\usepackage{color}
\definecolor{orange}{rgb}{1,0.5,0}
\definecolor{brown}{rgb}{0.65, 0.16, 0.16}
\definecolor{phlox}{rgb}{0.87, 0.0, 1.0}

\graphicspath{{figs/}}                  

\begin{document}

    \title{Generation of circular polarization of CMB due to the
Euler-Heisenberg Lagrangian }
\author{Iman Motie}
\email{i.motie@alumni.iut.ac.ir}
\affiliation{ Department of Physics, Mashhad Branch, Islamic Azad University, Mashhad, Iran }

    \begin{abstract}
        It is known that the dominant contribution of primordial anisotropies of Cosmic Microwave Background (CMB) is linearly polarized via Compton scattering.
However circular polarization of this anisotropy has not been observed up to now, but
it has not been excluded in observational evidence. Here we show that the Euler-Heisenberg Effective Lagrangian can generate circular polarization in CMB. This
generation is calculated by using the Quantum Boltzmann Equation for the time evolution
of the Stokes parameter. Also, we estimate the Faraday conversion phase in order of $10^{-13}
$
due to this Lagrangian which gives a lower bound on circular polarization of CMB
.
     
    \end{abstract}

    \maketitle

\section{Introduction}

Different interactions of photons can alter their polarization states as Faraday rotation, Faraday conversion, and so on. In some special cases measurement of circular
polarization contribution is a very important tool to understand the universe. In standard Cosmology, CMB photons are assumed to be linearly polarized partially. In other words, the quota of circular polarization is considered to be ignorable, because there are no
generating mechanisms in the epoch of recombination \cite{yek, 2}. Theoretical studies of
the CMB polarization predict that $10\%$ of the CMB anisotropies are linearly polarized
\cite{3,4}. There is consistency between observational results and theatrical predictions
\cite{5,6,7,8,9} However, the generation of the circular polarization of the CMB can occur in
the free-streaming regime after the epoch of recombination. It is shown that Compton
(Thomson) scattering cannot generate the circular polarization \cite{10}. There are some
mechanisms, which can enhance the generation of circular polarization. The role of the
Lorentz violation in the extension of the standard model (SME), noncommutative QED
(NC), background magnetic field, gravitation effects, and axion-like cosmology to generate
the circular polarization of the CMB are shown in refs \cite{11,12,13,14,15,16,17,18}.
To describe polarization of radiation fields, one uses the Stokes parameters
$I$, $Q$, $U$ and $V$, where $I$ indicates the intensity of radiation field, $Q$
and $U$
describe linear polarization intensities and $V$
denotes circular polarization intensity
of radiation fields \cite{19}. The time evolution of these parameters is given through the
Quantum Boltzmann equation. The Boltzmann equation is a systematic mechanism
that describes the evolution of the distribution function under collisions and gravity.
In principle, linear polarizations of the CMB radiation field propagating from the last
scattering surface can rotate each other and convert to circular polarizations by effects
of background fields, particle scattering, and temperature fluctuations. The conversion
of linear to circular polarization is described by the formalism of Faraday conversion
(FC) \cite{20}. The conversion from linear to circular polarizations is explained by the time
evolution of the Stokes parameter $V$,
\begin{eqnarray} \label{I1}
&\dot{V}= 2
U\frac{d}{dt}(\Delta\Phi_{FC}),
\end{eqnarray}
where $\Delta\Phi_{FC}$ is the Faraday conversion phase shift \cite{11}, quantitatively determining the
conversion from linear polarizations to circular polarizations.
In Ref. \cite{12} is shown that the FC phase shift which is generated by primordial
magnetic field and NC is in order of magnitude  $10^
{-19}$ and $10^
{-17}$ respectively. Ref.\cite{17}
shows that the FC phase shift by considering the nonlinear photon-photon interaction is about $10^{-16}$.

In this article, we consider the effective Euler-Heisenberg Lagrangian \cite{21,22}, and
by using the Quantum Boltzmann Equation (QBE)( see Ref. \cite{10} and the references
therein), we study the time-evolution of the Stokes parameter
$V$ , which quantitatively
determines circular polarizations. We obtain the CMB circular polarization intensity $V/T\sim5.2\times10^{-12}\Delta T/T $ and the corresponding FC phase shift $\Delta\Phi_{FC}\simeq6.2\times 10^{-13}$.
At the end, we discuss this circular polarization effect in an
X-ray laser experiment.

\section{ Polarization and Stokes parameters}
\label{sec2}

Let's consider quasi-monochromatic electromagnetic waves propagating in the ˆz-direction, in which the electric and magnetic fields vibrate on the x-y plane
\begin{eqnarray} \label{I2}
E_x = a_x(t) \cos[\omega_0t- \theta_x(t)],\,\,\,\, E_y = a_y(t) \cos[\omega_0t -\theta_y
(t)],
\end{eqnarray}
where amplitudes
$a_{x,y}$ and phase angles
$\theta_{x,y}$ are slowly varying functions with respect
to the frequency $\omega_0$. Any correlation between the $a_{x}$
and $a_{y}$-components indicates
polarizations of electromagnetic waves.
For a description of the polarization states of a nearly monochromatic electromagnetic
wave, one uses the Stokes parameters of radiation, $I$, $Q$, $U$ and $V$. Where the
parameter  $I$ is total intensity,  $Q$ and  $U$ are intensities of linear polarizations of
electromagnetic waves and the $V$ parameter indicates the difference between left- and
right- circular polarization intensities. Linear polarization can also be characterized
through a vector of modulus $P_L\equiv \sqrt{Q^2+U^2}
$
\cite{19}.
In the quantum mechanic’s description, an arbitrary polarized state of a photon $
(|k^
0|^2
= |\bf k|^2
)$ propagating in the $\hat{z}$-direction can also be written as
\begin{eqnarray}
|\epsilon\rangle=a_1\exp(i\theta_1)|\epsilon_1\rangle+a_2\exp(i\theta_2)|\epsilon_2\rangle,
\end{eqnarray}
where linear bases $|\epsilon_1\rangle$ and
$|\epsilon_2\rangle$ indicate the polarization states in the $x$-
and $y$-directions 
and the Stokes parameters in this description are given by \cite{10}
\begin{eqnarray}
\hat{I}&=&|\epsilon_1\rangle\langle\epsilon_1|+|\epsilon_2\rangle\langle\epsilon_2|,\nonumber\\
\hat{Q}&=&|\epsilon_1\rangle\langle\epsilon_1|-|\epsilon_2\rangle\langle\epsilon_2|,\nonumber\\
\hat{U}&=&|\epsilon_1\rangle\langle\epsilon_2|+|\epsilon_2\rangle\langle\epsilon_1|,\nonumber\\
\hat{V}&=&i|\epsilon_2\rangle\langle\epsilon_1|-i|\epsilon_1\rangle\langle\epsilon_2|.
\label{i-v}
\end{eqnarray}
An ensemble of photons in a general mixed state is described by a normalized density matrix $\rho_{ij}\equiv
(\,|\epsilon_i\rangle\langle \epsilon_j|/{\rm tr}\rho)$, and the dimensionless expectation values for Stokes
parameters are given by
\begin{eqnarray}
   I\equiv\langle  \hat I \rangle &=& {\rm tr}\rho\hat I
=1,\label{i}\\
Q\equiv\langle  \hat Q \rangle &=& {\rm tr}\rho\hat{Q}=\rho_{11}-\rho_{22},\label{q}\\
U\equiv\langle
 \hat U\rangle &=&{\rm tr}\rho\hat{U}=\rho_{12}+\rho_{21},\label{u}\\
V\equiv\langle  \hat V \rangle &=& {\rm
tr}\rho\hat{V}=i\rho_{21}-i\rho_{21}, \label{v}
\label{ic}
\end{eqnarray} 
where “tr” indicates the trace in the space of the polarization states. These above
equations determine the relationship between four Stokes parameters and the $2\times2$
density matrix $\rho$ of photon polarization states
\begin{eqnarray}
    \rho =\frac{1}{2}\begin{pmatrix}I+Q & U-iV\\ U+iV & I-Q\end{pmatrix}
\end{eqnarray}
The time evolution of these Stokes parameters is given through the Boltzmann
equation. Ones can consider each polarization state of the CMB radiation as the phase
space distribution function $\xi$. The classical Boltzmann equation generally is written as
\begin{eqnarray}
    \frac{d}{dt}\xi=\mathcal{C}(\xi),
\end{eqnarray}
where the left-hand side is known as the Liouville term and the right-hand side of the
Boltzmann equation contains all possible collision terms. By considering the photon
interaction on the right-hand side of the above equation, we can calculate the time evolution
of each polarization state of the photons. In the next section, we consider the
photon-photon scattering via the Euler-Hesinberg Lagrangian to computation of the
polarization’s states. 

\section{The Euler-Heisenberg Lagrangian and the photons polarizatio}

Lets to consider the free gauge field
$A_\mu$ in terms of plane wave solutions in the Coulomb
gauge [26]
\begin{eqnarray}
 A_\mu(x) = \int \frac{d^3 k}{(2\pi)^3
2 k^0} \left[ a_i(k) \epsilon _{i\mu}(k)
        e^{-ik\cdot x}+ a_i^\dagger (k) \epsilon^* _{i\mu}(k)e^{ik\cdot x}
        \right],
\end{eqnarray}
where $ \epsilon _{i\mu}(k)$
 are the polarization four vectors and the index
$i = 1, 2$, representing two
transverse polarizations of a free photon with four-momentum
$k$ and $k_0= |\bf k|$. In addition $a_i(k)$  and $ a_i^\dagger (k)$
are creation and annihilation operators, which satisfy the canonical commutation relation
\begin{equation}
        \left[  a_i (k), a_j^\dagger (k')\right] = (2\pi )^3 2k^0\delta_{ij}\delta^{(3)}({\bf k} - {\bf k}' ).
\label{comm}
\end{equation}
The density operator, describing a system of photons is given by
\begin{equation}
\hat\rho=\frac{1}{\rm {tr}(\hat \rho)}\int\frac{d^3p}{(2\pi)^3}
\rho_{ij}(p)a^\dagger_i(p)a_j(p),
\end{equation}
where $\rho_{ij}(k)$ is the general density-matrix in the space of polarization states with a fixed energy-momentum $k$
, i.e. Eq. (5-8), and $a^\dagger_i(k)a_j(k)$ is the number operator ($D^0_{ij}(k)$), which is related to the density matrix  $\rho_{ij}(k)$ as
\begin{eqnarray}
\langle\, D^0_{ij}(k)\,\rangle\equiv {\rm tr}[\hat\rho
D^0_{ij}(k)]=(2\pi)^3 \delta^3(0)(2k^0)\rho_{ij}(k).    
\end{eqnarray}
On the other hand, by considering the Heisenberg picture, the time evolution of the operator $D^0_{ij}(k)$, is
\begin{eqnarray}
\frac{d}{dt}D^0_{ij}(k) = i\left[H^0_I
(t);D^0_{ij}(k)\right],
\end{eqnarray}
where $H$ is the full Hamiltonian. The expectation value of this equation gives the Boltzmann equation (10) for the density matrix of the system (as well as the polarization
states), which is a generalization of the usual classical Boltzmann equation for particle occupation numbers. By substituting Eq. (14) in Eq. (15), the time evolution of $\rho_{ij}(k)$ as well as Stokes parameters is given by \cite{10},
\begin{eqnarray}
 (2\pi)^3 \delta^3(0)(2k^0)
\frac{d}{dt}\rho_{ij}(k) = i\langle \left[H^0_I
(t);D^0_{ij}(k)\right]\rangle-\frac{1}{2}\int dt\langle
\left[H^0_I(t);\left[H^0_I
(0);D^0_{ij}(k)\right]\right]\label{bo}\rangle,   
\end{eqnarray}
where
 $H^0_I(t)$ is the first order of the interacting Hamiltonian. The first term on the right-handed side is a forward scattering term, and the second one is a higher-order
collision term. For computation of the expectation values, one can use [10],  
\begin{eqnarray}
\langle \, a^\dag_{s'}(p')a_{s}(p)\, \rangle
&=&2p^0(2\pi)^3\delta^3(\mathbf{p}-\mathbf{p'})\rho_{ll'}(p),\nonumber\\
\langle p|\, a^\dag_{s'}(p')a_{s}(p)a^\dag_{l'}(q')a_{l}(q)\,|p \rangle
\!\!\!&=& \!\!\!\langle p|\,
a^\dag_{s'}(p')a^\dag_{l'}(q')a_{s}(p)a_{l}(q)\,|p \rangle\nonumber\\
&+&\!\!\!2p^0(2\pi)^3\delta^{sl'}\delta^3(\mathbf{p}-\mathbf{q'})\langle p|\, a^\dag_{s'}(p')a_{l}(q)\,|p \rangle\nonumber\\
&=&\!\!\!\ 4p^0q^0(2\pi)^6\delta^3(\mathbf{p}-\mathbf{p'})
\delta^3(\mathbf{q}-\mathbf{q'})\rho_{ss'}(p)\rho_{ll'}(q)\nonumber\\
&+&\!\!\!
4p^0q^0(2\pi)^6\delta^3(\mathbf{p}-\mathbf{q'})
\delta^3(\mathbf{q}-\mathbf{p'})\rho_{s'l}(q)[\delta_{sl'}+\rho_{sl'}(p)],   
\end{eqnarray}
where at first applying the Wick theorem to arrange all creation operators to the left and all annihilation operators to the right, then use the contraction rule (17) of operators $a_i^\dag$ and $a_j$.

At now, to the computation of the photon polarization states we consider the photon-photon scattering via the Euler-Heisenberg Lagrangian. The Euler-Heisenberg lagrangian
is a low-energy effective lagrangian describing multiple photon interactions. The
Euler-Heisenberg lagrangian is the first correction to the Maxwell lagrangian for weak
fields ($E, B\ll m^2/e$, $m$ is the electron mass), and it is given by \cite{21,22}
\begin{eqnarray}
  \frac{2\alpha^2}{45m^4}(4\mathcal{F}^2+7\mathcal{G}^2),  \label{eh}
\end{eqnarray}
where
\begin{eqnarray}
 \mathcal{F}=-\frac{1}{4}F_{\mu\nu}F^{\mu\nu}, \;\;\mathcal{G}=-\frac{1}{4}F_{\mu\nu}\tilde F^{\mu\nu},  \label{eh}
\end{eqnarray}
where  $F_{\mu\nu}=\partial_\mu A_\nu-\partial_\nu A_\mu$ is the electromagnetic field strength and $\tilde{F}^{\mu\nu}=\epsilon^{\mu\nu\alpha\beta}F_{\alpha\beta}$ in which, $\epsilon^{\mu\nu\alpha\beta}$
is an anti-symmetric tensor of rank four (See review articles \cite{23} and \cite{24}).
In the \cite{17}, we considered the role of $(F_{\mu\nu}\tilde F^{\mu\nu})^2$ term in Eq. (19) in the generation of
the circular polarization.

In this work, we study both terms in (19) by using the Stokes parameters approach
and QBE (16), we compute their role of them in the generation of circular polarization.

It is worthwhile to mention that Kosowsky has shown that Maxwell Lagrangian $F_{\mu\nu} F^{\mu\nu})$ has no any contribution at the first order of QBE (The first term in the RHS
of Eq. (16) is zero) and its contribution comes from the second order \cite{10}.
The effective Euler-Heisenberg Lagrangian (19), corresponds to the interacting Hamiltonian
\begin{eqnarray}
 \delta \pounds &\approx &
-\frac{\alpha^2}{90m^4}
\left[(F_{\mu\nu}F^{\mu\nu})^2
+\frac{7}{4}(F_{\mu\nu}\tilde{F}^{\mu\nu}
)^2\right].
\end{eqnarray}
which is perturbatively small, at order of $\alpha^2$, so that we only compute the first order of Quantum Boltzmann Equation i.e. the first term in RHS of the Eq. (16), and neglect
the second term which is of order of $\alpha^4$. In principle when first term does not have any
result, in special theory, people try to compute the second term.

As a result, the time-evolution of the density matrix approximately obtain as
\begin{eqnarray}
   (2\pi)^3 \delta^3(0)2k^0
\frac{d}{dt}\rho_{ij}(k) \!\!&\approx& i\langle
\left[H^0_I(t),D^0_{ij}(k)\right]\rangle\nonumber\\
&=&-\frac{2i\alpha^2}{45m^4}(2\pi)^3\delta^3(0)\! \int\frac{d^3p}{(2\pi)^32p^0}\Big[ (p.k)^2[\epsilon_{s}(k).\epsilon_{s'}(p)\epsilon_{l}(k).\epsilon_{l'}(p)]\nonumber\\
&\times&[-5\rho_{s'l'}(p)\rho_{is}(k)\delta^{lj}+5\rho_{s'l'}(p)\rho_{lj}(k)\delta^{si}+4\rho_{l's'}(p)\rho_{lj}(k)\delta^{si}\nonumber\\
&-& 4\rho_{l's'}(p)\rho_{i s}(k)\delta^{lj}+3\rho_{l's'}(p)\rho_{sj}(k)\delta^{li}-3\rho_{l's'}(p)\rho_{il}(k)\delta^{sj}\nonumber\\
&+&4\rho_{s'l'}(p)\rho_{sj}(k)\delta^{li}-4\rho_{s'l'}(p)\rho_{il}(k)\delta^{sj}+9\rho_{lj}(k)\delta^{si}\delta^{s'l'}\nonumber\\
&-&9\rho_{is}(k)\delta^{lj}\delta^{s'l'}+3\rho_{sj}(k)\delta^{s'l'}\delta^{li}-3\rho_{il}(k)\delta^{sj}\delta^{s'l'}
]\nonumber\\
&+&[p.\epsilon_{s}(k)k.\epsilon_{s'}(p)p.\epsilon_{l}(k)k.\epsilon_{l'}(p)-
2(p.k)\epsilon_{s}(k).\epsilon_{s'}(p)p.\epsilon_{l}(k)k.\epsilon_{l'}(p)]\nonumber\\
&\times&[8\rho_{lj}(k)\delta^{si}\delta^{s'l'}-8\rho_{is}(k)\delta^{s'l'}\delta^{lj}+4\rho_{l's'}(p)\rho_{lj}(k)\delta^{si}
\nonumber\\
&-&4\rho_{l's'}(p)\rho_{is}(k)\delta^{lj}-4\rho_{s'l'}(p)\rho_{is}(k)\delta^{lj}+4\rho_{l's'}(p)\rho_{lj}(k)\delta^{si}
\nonumber\\
&+&4\rho_{sj}(k)\delta^{l's'}\delta^{li}+4\rho_{sj}(k)\rho_{l's'}(p)\delta^{li}-4\rho_{il}(k)\delta^{l's'}(k)\delta^{sj}
\nonumber\\
&-&4\rho_{l's'}(p)\rho_{il}(k)\delta^{sj}+4\rho_{sj}(k)\rho_{s'l'}(p)\delta^{li}-4\rho_{s'l'}(p)\rho_{il}(k)\delta^{sj}
]\nonumber\\
&-&28\epsilon^{\mu\nu\alpha\beta}\epsilon^{\sigma\nu'\gamma\beta'}k_\gamma k_\mu p_\alpha p_\sigma\epsilon_{s'\beta}(p)\epsilon_{l\nu'}(p)\epsilon_{s\nu}(k)\epsilon_{l'\beta'}(k)\nonumber\\
&\times&[\rho_{l'j}(k)\delta^{si}-\rho_{is}(k)\delta^{l'j}+\rho_{sj}(k)\delta^{l'i}-\rho_{il'}(k)\delta^{sj}]\nonumber\\
&\times&[\rho_{ls'}(p)+\rho_{s'l}(p)+\delta^{s'l}
]
\Big],
\label{j1} 
\end{eqnarray}
where $k$  and
$p$ indicate the energy-momentum states of photons and $\delta^3(0)$
 will be canceled in the final expression. For the rest of the paper, we do not consider the terms with
linear dependence of $\rho_{ij}$
 on the right side of the above equation, because we are interested
in photon-photon forward scattering. Using Eq. (21), the time-evolution of Stocks
parameters (5-8) are obtained as follows
\begin{eqnarray}
    \dot I (\bf K)=0.
\end{eqnarray}

 It implies, in the ensemble of photons, the total intensity in any direction $\hat k$ is
constant and does not change from Euler-Heisenberg's forward interaction. The above
result for intensity
$I$ is expected, because we study forward scattering which cannot change momenta of photons (necessary condition to change intensity in any direction). 
\begin{eqnarray}
\dot{Q}(k)
&=&\frac{16\alpha^2}{45m^4k^0}V(\mathbf k )\int\frac{d^3p}{(2\pi)^32p^0}(p^0k^0)^2\Big[f_1(\hat p, \hat k) U(\mathbf{p})-g_1(\hat p, \hat k)V(\mathbf p)\Big]
,\label{qd}\\
\dot{U}(k)
&=&\frac{8\alpha^2}{45m^4k^0}V(\mathbf k )
\int\frac{d^3p}{(2\pi)^32p^0}(p^0k^0)^2\Big[f_2(\hat p, \hat k) I(\mathbf{p})+f_3(\hat p, \hat k) Q(\mathbf{p})+ g_2(\hat p, \hat k)Q(\mathbf p)\Big]
,
\label{ud}\\
\dot{V}(k)
&=&\frac{8\alpha^2}{45m^4k^0}
\int\frac{d^3p}{(2\pi)^32p^0}(p^0k^0)^2\Big[f_4(\hat p, \hat k) I(\mathbf{p})U(\mathbf{k})+f_5(\hat p, \hat k) Q(\mathbf{p})U(\mathbf{k})\nonumber\\&+& ( f_6(\hat p, \hat k)+g_1(\hat p, \hat k)Q(\mathbf k)U(\mathbf{p})+
g_2(\hat p, \hat k)V(\mathbf k)Q(\mathbf{p})
\Big].
\label{vd} 
\end{eqnarray}
In above equations, we have two groups of functions,
$f_i(\hat p, \hat k)$ and  $g_i(\hat p, \hat k)$
(see details of these dimensionless functions in Appendix), which presents the contribution of
$\mathcal{F}$ and $\mathcal{G}$ terms of the effective Euler-Heisenberg interaction (19) respectively. As equations (23,24) show, the initial circular polarization of an ensemble of photon V (k)
can be converted to a linear one $U(\bf k), Q(\bf k)$
due to Euler-Hisenberg interactions. Note $I(\bf p), U(\bf p), Q(\bf p) $ and $V(\bf p)$ are Stokes parameters that describe the target photons.
Another condition that is needed to convert circular polarization to linear one by
Euler-Heisenberg interactions are that $I(\bf p), U(\bf p), Q(\bf p) $ and $V(\bf p)$ 
must have some degrees
of anisotropies [see right side of equations (23,24)]. Let’s discuss the time-evolution of
circular polarization
$\dot V$ given in (25) in more detail. As shown in (25), $\dot V(\bf k)$
 is proportional to $U(\bf k)$ and  $Q(\bf k)$ modes. This indicates that an ensemble of linearly polarized
photons will acquire circular polarizations. This conversion from linear polarized modes
to circular polarized modes is also described by Faraday conversion phase shift $\Delta\phi _{FC}$
obeying Eq. (1).

\section{$\Delta\phi _{FC}$ due to CMB-CMB Forward Scattering}

Using Eqs. (1) and (25), we have,
\begin{eqnarray}
 \Delta\phi_{FC} &\simeq &
\frac{2\alpha^2}{45}\int dt\int\frac{p^0d^3p}{(2\pi)^3}\frac{k^0}{m_e^4}\Big[f_4(\hat p, \hat k) I(\mathbf{p})+f_5(\hat p, \hat k)Q(\mathbf p)+ (f_6(\hat p, \hat k)+ g_1(\hat p, \hat k))U(\mathbf p)\Big].
\label{result2}
\end{eqnarray}
To estimate the evaluation of $\dot V$
 as well as $\Delta \phi_{FC}$, first we transform the integral over into the integral over Redshift as follows
\begin{eqnarray}
\int dt'\rightarrow \frac{1}{H_0}\int \frac{dz'}{(1+z')\hat H(z')},  
\end{eqnarray}
where $\hat H(z)$ is 
\begin{eqnarray}
    \hat H(z)=[\Omega_r(1+z)^4+\Omega_M(1+z)^3+\Omega_\Lambda)]^{1/2},
\end{eqnarray}
 where $H_0$ is the Hubble parameter, $\Omega_r\leq10^{-4}$, 
$\Omega_M\simeq 0.3$ and $\Omega_\Lambda\simeq 0.7$ are present
densities of radiation, matter, and dark energy, respectively. Note energy and intensity
of photons depend on the Redshift as cosmic expansion results in the universe,
\begin{eqnarray}
    E=E_0(1+z),\:\:  I=I_0(1+z)^4,\;\;Q=Q_0(1+z)^4,\;\; U=U_0(1+z)^4,\;\;
\end{eqnarray}
where $E_0,I_0, Q_0$ and $U_0$ 
are measured at the present time. And also we use
\begin{eqnarray}
    \int\frac{p^0d^3p}{(2\pi)^2}I(\mathbf{p})=\bar I\approx\bar p\,n|_{CMB}.
\end{eqnarray}
where $\bar I$
 is total intensity of CMB photon, $\bar p=T_{CMB}$  is average energy of CMB and $n|_{CMB}$
is the number density of CMB photons. Using Eqs.(1)-(30) and integrating from
last scattering Redshift up to the present time, the Faraday conversion phase shift can be
written as following 
\begin{eqnarray}
\Delta\phi_{FC} &\simeq &
\Big(\frac{2\alpha^2}{45m^4}\Big)\bar I_0k^0H_0^{-1}\Gamma\int^{LSS}_{0}dz\frac{(1+z)^4}{[\Omega_r(1+z)^4+\Omega_M(1+z)^3+\Omega_\Lambda)]^{1/2}},
\end{eqnarray}
where 
\begin{eqnarray}
    \Gamma=\frac{1}{\bar I}\int \frac{d\Omega}{4\pi}\Big[f_4(\hat p, \hat k) I(\mathbf{p})+f_5(\hat p, \hat k)Q(\mathbf p)+ (f_6(\hat p, \hat k)+ g_1(\hat p, \hat k))U(\mathbf p)\Big].
\end{eqnarray}
where $\bar I_0$
is the average intensity of CMB at the present time and
$z_{LSS}
\simeq 1100$ is last
scattering surface redshift. Note $\Gamma$ includes the CMB’s anisotropies in temperature (first
term in above equation) and polarization (second and third terms). We just consider $\Gamma\approx\frac{\delta T}{T}\simeq 10^{-5}$ 
which is in order of the CMB’s anisotropies. Considering all of the discussions
in the present chapter, we have
\begin{eqnarray}
    \Delta\phi_{{FC}|_{CG} }&\simeq &10^{-6}rad 
\frac{k_0}{2.3\times10^{-4} eV}\frac{\bar p_0}{2.3\times10^{-4} eV}\frac{n|_{CMB}(z=0)}{411 cm^{-3}}\frac{\Gamma}{10^{-5}}.
\end{eqnarray}

It should be mentioned here that same calculation is done to obtain $\Delta\phi_{FC}$ for gamma
ray bursts (GRB) due to CMB-GRB forward scattering [25].

\section{Conclusion and remarks}
In this article, we approximately solve the first order of the Quantum Boltzmann Equation for the density matrix of photon ensemble, and the time-evolution of Stokes parameters,
and we show the effect that propagating photons convert their linear polarizations to
circular polarizations by the nonlinear Euler-Heisenberg effective Lagrangian in QED.
Applying CBM photons, we obtain the circular polarization intensity $(V/T)\simeq 5.2 \times 10{-12}(\Delta T/T)$, twelve orders of magnitude smaller than the CMB anisotropy
$(\Delta T/T)$, corresponding to a Faraday conversion phase shift $\Delta \Phi_{FC}\simeq6.2\times 10^{-13}$. This
result gives a low bound of CMB circular polarizations due to the nonlinear Euler-Heisenberg effective interaction. To end this article, we remark on this circular polarization effect in an
X-ray laser experiment. Our results (22-25) imply that a linear polarized laser beam must develop circular polarizations while it is propagating. In this
case, the amplitudes of linear and circular polarizations are comparable, one needs to
completely solve Eqs. (22-25) to respectively obtain the intensities of linear and circular
polarizations of the laser beam.

\section{Acknowledgment}
I would like to thank R. Mohammadi and S. S. Xue for fruitful discussion

\appendix
\section{{Details of the computations}}
\label{appe}
For details of the calculation please have a look at the Appendix of the published version.



\begin{thebibliography}{99}
\bibitem{yek}
 R. B. Partridge, J. Nowakowski and H. M. Martin, Nature 331, 146 (1988).

\bibitem{2}
G. F. Smooth, et al. Astrophysical Journal Letters 396 (1), L1 (1992); C. L .Bennett, et al. Astrophysical Journal Letters 464, L1 (1996).


\bibitem{3} R. Crittenden, R. Davis and P. Steinhardt, Ap. J. 417, L13 (1993).

\bibitem{4}
 R. A. Frewin, A. G. Polnarev and P. Coles, Mon. Not. R. Ast. Soc. 266, L21 (1994);
D. Harari and M. Zaldarriaga, Phys. Lett. B 319, 96 (1993).


\bibitem{5}
Kovac, J.M. et al. (2002). ”Detection of polarization in the cosmic microwave background using DASI”. Nature 420 (6917): 772787.

\bibitem{6} A. D. Miller, et al. Astrophysical Journal 521 (2), L79 (1999).

\bibitem{7}
 A. Melchiorri, et al. Astrophysical Journal 536 (2), L63 (2000).
 
 \bibitem{8}
 S. Hanany, et al. Astrophysical Journal Letters 545 (1), L5 (2000).
 
\bibitem{9}
G. Hinshaw, et al. Astrophysical Journal (Supplement Series) 170(2), 288 (2007).

 
\bibitem{10}
A. Kosowsky, Annals Phys. 246, 49 (1996) [arXiv:astro-ph/9501045].

\bibitem{11}
A. Cooray, A. Melchiorri and J. Silk, Phys. Lett. B 554, 1 (2003) [arXiv:astroph/0205214].

\bibitem{12} 
M. Zarei, E. Bavarsad, M. Haghighat, R. Mohammadi, I. Motie, Z. Rezaei, Phys. Rev. D 81, 084035 (2010) [arXiv:hep-th/0912.2993].

\bibitem{13}
M. Giovannini, [arXiv:hep-ph/0208152] (2002).

\bibitem{14}
M. Giovannini and K. E. Kunze, Phys. Rev. D 78, 023010 (2008) [arXiv:astroph/0804.3380].

\bibitem{15}
S. Alexander, J. Ochoa and A. Kosowsky, Phys. Rev. D 79, 063524 [arXiv:astroph/0810.2355] (2009).

\bibitem{16}
F. Finelli and M. Galaverni, Phys. Rev. D 79, 063002 [arXiv:astro-ph/0802.4210]
(2009).

\bibitem{17} 
I. Motie and S. S. Xue, EPL 100, 17006 (2012).

\bibitem{18}
R. Mohammadi, “Evidence for cosmic neutrino background form CMB circular polarization,” Eur. Phys. J. C 74:3102(2014), arXiv:1312.2199 [astro-ph.CO];
J. Khodagholizadeh, R. Mohammadi and S-S. Xue, Phys. Rev. D 90, 091301
(2014),[arXiv:1406.6213[astro-ph.CO]]; R. Mohammadi, J. Khodagholizadeh,
M. Sadegh and S. S. Xue, [arXiv:1602.00237 [astro-ph.CO]].

\bibitem{19}
J. D. Jackson, Classical Electrodynamic, Wiley and Sons: New York (1998).

 \bibitem{20}
 T. W. Jones and S. L. ODell, Astrophys J. 214, 522 (1977); M. Ruszkowski and
M. C. Begelman, [arXiv:astro-ph/0112090] (2001).

\bibitem{21}
W. Heisenberg and H. Euler, Flogerungen aus der Diracschen Theorie des Positron,
Z. Phys. 98 (1936) 714; English translation: physics/0605038. H. Euler, Ann. d.
Phys. 26 (1936) 398; J. Schwinger, Phys. Rev. 82 (1951) 664.

\bibitem{22}
V. Weisskopf, “The electrodynamics of the vacuum based on the quantum theory
of the electron”, Kong. Dans. Vid. Selsk. Math-fys. Medd. XIV No. 6 (1936);

\bibitem{23}
R. Ruffini, G. Vershchagin, S. S. Xue, Physical Reports 487, 1-140 (2010).

 \bibitem{24}
 G. V. Dunne, [arXiv:hep-th/0406216] (2004).


\bibitem{25}
S. Batebi, R. Mohammadi, R. Ruffini, S. Tizchang, S.S. Xue, Phys. Rev. D 94
(2016) 065033.

\bibitem{26}
C. Itzykson, J. B. Zuber: Quantum field theory, McGraw-Hill: United States of
America (1980).

 \bibitem{27}
 D. J. Fixsen, et al. Astrophysical Journal 473, 576 (1996).


\bibitem{28}
E. L. Wright, [arXiv:astro-ph/0305591] (2003).

\bibitem{29}
G. L. Kotkin, V. G. Serbo, [arXiv:hep-ph/9611345] (1996).
\end{thebibliography}
\end{document}